\documentclass[12pt,a4paper,twoside]{article}
\usepackage{a4}
\usepackage{amsmath}
\usepackage{amssymb}
\usepackage{array}
\IfFileExists{bbm.sty}{\usepackage{bbm}}{\typeout{ * ^^J
   * Please install the bbm bundle! I try to replace the bbm fonts by the ^^J
   * AMS blackboard font and the cm bold font ^^J *}
   \newcommand{\mathbbm}[1]{\mathbb{##1}}}
\IfFileExists{cite.sty}{\usepackage{cite}}{}

\makeatletter
\def\preprint#1{\def\@preprint{#1}} \def\@preprint{}
\def\email#1{\def\@email{#1}} \def\@email{}
\def\address#1{\def\@address{#1}} \def\@address{} 
\def\@maketitle{%
  \newpage
  \null
  {}\hfill \raisebox{3cm}{\if \@preprint \empty \else 
			  {\renewcommand{\arraystretch}{1}
			  \begin{tabular}{c}\@preprint\end{tabular}} \fi}
  \vspace{-2cm}
  \begin{center}%
  \let \footnote \thanks
    {\LARGE\sffamily \@title \par}%
    \vskip 1.5em%
    {\large\scshape
      \lineskip .5em%
      \begin{tabular}[t]{c}%
        \@author
      \end{tabular}\par}%
    \vskip 1em%
    {\if \@address\empty \else \large\slshape
      \lineskip .5em%
      \begin{tabular}[t]{c}%
	\@address
      \end{tabular}\par \fi}%
     \vskip 0.5em%
    {\if \@email\empty \else \large e-mail:\space %
				   \texttt{\ignorespaces \@email} \fi}
     \vskip 1.5em%
    {\large \@date}%
  \end{center}%
  \par
  \vskip 2em}
\makeatother

\makeatletter
\renewenvironment{thebibliography}[1]{%
  \@xp\section\@xp*\@xp{\refname}%
  \normalfont\footnotesize\labelsep.5em\relax
  \renewcommand\theenumiv{\arabic{enumiv}}\let\p@enumiv\@empty
  \list{\@biblabel{\theenumiv}}{\settowidth\labelwidth{\@biblabel{#1}}%
    \parskip 0pt \itemsep 0pt 
    \leftmargin\labelwidth \advance\leftmargin\labelsep
    \usecounter{enumiv}}%
  \sloppy \clubpenalty\@M \widowpenalty\clubpenalty
  \parskip 0pt \itemsep 0pt \parsep 0pt \partopsep 0pt 
  \sfcode`\.=\@m
}{%
  \def\@noitemerr{\@latex@warning{Empty `thebibliography' environment}}%
  \endlist
}
\makeatother

\makeatletter
\def\eqnarr@left{\@centering}
\let\eqnarr@opts\relax
\def\equationarray{%
 \col@sep\arraycolsep
 \def\d@llarbegin{$\displaystyle}%
 \def\d@llarend{$}%
 \stepcounter{equation}%
 \let\@currentlabel=\theequation
 \set@eqnsw \global\@eqcnt\z@ \global\@eqargcnt\z@
 \let\@classz\@eqnclassz
 \def\multicolumn##1##2##3{\@eqnmulticolumn{##1}{##2}{##3}%
                           \global\advance\@eqcnt##1
                           \global\advance\@eqcnt\m@ne}%
 \def\@halignto{to\displaywidth}%
 \@ifnextchar[{\@equationarray}{\@equationarray[.]}}
\let\@eqnmulticolumn=\multicolumn
\def\yesnumber{\global\@eqnswtrue}
\let\set@eqnsw=\yesnumber
\def\@amper{&}
\newcount\@eqargcnt  % counts number of columns
\def\@equationarray[#1]#2{%
     \eqnarr@opts
     \@tempdima \ht \strutbox
     \advance \@tempdima by\extrarowheight
     \setbox\@arstrutbox=\hbox{\vrule
               \@height\arraystretch \@tempdima
               \@depth\arraystretch \dp \strutbox
               \@width\z@}%
     \gdef\advance@eqargcnt{\global\advance\@eqargcnt\@ne}%
     \begingroup
     \@mkpream{#2}%
     \xdef\@preamble{%
      \if #1l\tabskip\z@ \else\if #1r\tabskip\@centering
                         \else\if #1c\tabskip\@centering
                         \else\tabskip\eqnarr@left \fi\fi\fi
      \halign \@halignto
      \bgroup \tabskip\z@ \@arstrut \@preamble
      \if #1l\tabskip\@centering \else\if #1r\tabskip\z@
                                 \else\tabskip\@centering \fi\fi
      \@amper\llap{\@sharp}\tabskip\z@\cr}%
     \endgroup
     \gdef\advance@eqargcnt{}%
     \bgroup
     \let\@sharp## \let\protect\relax
     \m@th   \let\\=\@equationcr
     \let\par\@empty
     $$                            % $$ BRACE MATCHING HACK
     \lineskip \z@
     \baselineskip \z@
     \@preamble}
\def\@eqnclassz{\@classx
   \@tempcnta \count@
   \advance@eqargcnt
   \prepnext@tok
   \@addtopreamble{%
      \global\advance\@eqcnt\@ne
      \ifcase \@chnum
      \hfil \d@llarbegin \insert@column \d@llarend\hfil \or
      \d@llarbegin \insert@column \d@llarend \hfil \or
      \hfil\kern\z@ \d@llarbegin \insert@column \d@llarend \or
      $\vcenter
      \@startpbox{\@nextchar}\insert@column \@endpbox $\or
      \vtop \@startpbox{\@nextchar}\insert@column \@endpbox \or
      \vbox \@startpbox{\@nextchar}\insert@column \@endpbox
      \fi}\prepnext@tok}
\def\endequationarray{\@zequationcr
   \egroup
   \global\advance\c@equation\m@ne $$  % $$ BRACE MATCHING HACK
   \egroup\global\@ignoretrue
   \gdef\@preamble{}}
\def\@equationcr{${\ifnum0=`}\fi\@ifstar{\global\@eqpen\@M
    \@xequationcr}{\global\@eqpen\interdisplaylinepenalty
                   \@xequationcr}}
\def\@xequationcr{%
    \@ifnextchar[{\@argequationcr}{\ifnum0=`{\fi}${}%
    \@zequationcr}}
\def\@argequationcr[#1]{\ifnum0=`{\fi}${}\ifdim #1>\z@
   \@xargequationcr{#1}\else
   \@yargequationcr{#1}\fi}
\def\@xargequationcr#1{\unskip
   \@tempdima #1\advance\@tempdima \dp \@arstrutbox
   \vrule \@depth\@tempdima \@width\z@
   \@zequationcr\noalign{\penalty\@eqpen}}
\def\@yargequationcr#1{%
   \@zequationcr\noalign{\penalty\@eqpen\vskip #1}}
\def\@zequationcr{\@whilenum\@eqcnt <\@eqargcnt
   \do{\@amper\omit\global\advance\@eqcnt\@ne}%
   \@amper
   \if@eqnsw\@eqnnum\stepcounter{equation}\fi
   \set@eqnsw\global\@eqcnt\z@\cr}
\@namedef{equationarray*}{%
   \let\set@eqnsw=\nonumber \equationarray}
\@namedef{endequationarray*}{\endequationarray}
\makeatother

\DeclareMathOperator{\id}{id}
\DeclareMathOperator{\End}{End}
\DeclareMathOperator{\Hom}{Hom}
\DeclareMathOperator{\tr}{tr}
\DeclareMathOperator{\Tr}{Tr}
\DeclareMathOperator{\diag}{diag}

\def\ot{\otimes}
\def\op{\oplus}
\def\t{\tilde}
\def\h{\hat}
\def\iu{\mathrm{i}}

\def\bfd{\mathbf{d}}
\def\sfD{\mathsf{D}}

\def\bsj{\boldsymbol{\psi}}

\def\one{\mathbbm{1}}
\def\CX{{C^\infty (X)}}
\def\B{\mathcal{B}}
\def\C{\mathbbm{C}}
\def\N{\mathbbm{N}}
\def\M{\mathcal{M}}
\def\R{\mathbbm{R}}
\def\Z{\mathbbm{Z}}

\def\SU#1{\mathrm{SU(#1)}}
\def\U#1{\mathrm{U(#1)}}
\def\su#1{\mathrm{su(#1)}}
\def\u#1{\mathrm{u(#1)}}
\def\ul#1{\underline{#1}}

\def\itJ{\varPsi}		  \def\ittY{\tilde{\varUpsilon}}  
\def\itF{\varPhi}                 \def\ittJ{\tilde{\varPsi}}      
\def\itY{\varUpsilon}             \def\ittF{\tilde{\varPhi}}      
\def\itX{\varXi}                  \def\ittX{\tilde{\varXi}}      
\def\Mu{M_{\t{u}}}		  \def\Mn{M_{\t{n}}}

\def\yn{\yesnumber}
\def\npb{\nopagebreak}
\newcommand\eq[2][]{\begin{equation} \label{#1} #2 \end{equation}}
\newcommand\eqa[2]{\begin{equationarray*}{#1} #2 \end{equationarray*}}
\newcommand\eqas[3][]{\begin{equation}\label{#1} \begin{array}{#2} #3 
		      \end{array} \end{equation}}
\newcommand{\mb}[3][\!\!\!]{\left( #1 \begin{array}{#2} #3 
			    \end{array} #1 \right)} 
 
\def\mc{\multicolumn}
\newcommand{\ru}[2]{\vrule width0em height#1ex depth#2ex} 
\newcolumntype{Z}[1]{>{$\hfil}m{#1mm}<{$\hfil}>{$\hfil}m{#1mm}<{$\hfil}}
\newcolumntype{E}[1]{>{$\hfil}m{#1mm}<{$\hfil}}

\newcommand\al[1]{\begin{align} #1 \end{align}}

\newcommand{\W}[2][g]{\Omega^{#2}\mathfrak{#1}}
\renewcommand{\P}[2][g]{\pi(\Omega^{#2}\mathfrak{#1})}
\newcommand{\p}[2][a]{\hat{\pi}(\Omega^{#2}\mathfrak{#1})}
\newcommand{\D}[2][g]{\Omega_D^{#2} \mathfrak{#1}}
\newcommand{\J}[2][g]{\pi(\mathcal{J}^{#2}\!\mathfrak{#1})}
\newcommand{\cJ}[2][g]{\mathbbm{J}^{#2}\!\mathfrak{#1}}
\newcommand{\jj}[2][a]{\hat{\pi}(\mathcal{J}^{#2} \mathfrak{#1})}
\newcommand{\pf}[1][a]{\hat{\pi}(\mathfrak{#1})}
\newcommand{\hH}[2][g] {\hat{\mathcal{H}}^{#2} \mathfrak{#1}}

\renewcommand{\H}[2][g]{\mathcal{H}^{#2} \mathfrak{#1}}
\IfFileExists{bbm.sty}%
     {}%
     {}
\IfFileExists{bbm.sty}%
     {\newcommand{\cc}[2][a]{\mathbbm{c}^{##2}\!\mathfrak{##1}}}%
     {\newcommand{\cc}[2][a]{ \mathbf{c}^{##2}\!\mathfrak{##1}}}
\IfFileExists{bbm.sty}%
     {\newcommand{\tcc}[2][g]{\tilde{\mathbbm{c}}^{##2}\!\mathfrak{##1}}}%
     {\newcommand{\tcc}[2][g]{\tilde{ \mathbf{c}}^{##2}\!\mathfrak{##1}}}
\IfFileExists{bbm.sty}%
     {\newcommand{\bbr}[2][a]{\mathbbm{r}^{##2}\!\mathfrak{##1}}}%
     {\newcommand{\bbr}[2][a]{ \mathbf{r}^{##2}\!\mathfrak{##1}}}
\IfFileExists{bbm.sty}%
     {\def\one{\mathbbm{1}}}%
     {\def\one{\mathrm{I}}}

\def\mc{\multicolumn}
\newcommand{\f}[1][g]{\mathfrak{#1}}
\newcommand{\mat}[2][\C]{{\mathrm{M}}_{#2}{#1}}
\newcommand{\rf}[1][]{\textup{\eqref{#1}}}
\newcommand{\g}[1][5]{\gamma^{#1}}
\newcommand{\Ad}[1]{\mathrm{Ad}_{#1}\,}

\newtheorem{dfn}{Definition}

\def\vs{\vspace}
\def\th{\tfrac{1}{2}}

\def\tsum{\textstyle \sum}

\renewcommand{\arraystretch}{1.2}

\let\section=\subsection

\begin{document}

\title{     A Tour through Non--Associative Geometry}
\author{    Raimar Wulkenhaar}	
\address{   Institut f\"ur Theoretische Physik \\ 
	    Universit\"at Leipzig \\ 
	    Augustusplatz 10/11, D--04109 Leipzig, Germany}
\email{     raimar.wulkenhaar@itp.uni-leipzig.de} 
\preprint{  \texttt{hep-th/9607086}\\ revised version} 
\date{	    February 10, 1997}
\maketitle 

\vfill
\vfill

\begin{abstract}
\noindent
We develop a mathematical concept towards gauge field theories based upon 
a Hilbert space endowed with a representation of a skew--adjoint Lie 
algebra and an action of a generalized Dirac operator. This concept shares 
common features with the non--commutative geometry \`a la Connes\,/\,Lott, 
differs from that, however, by the implementation of skew--adjoint Lie 
algebras instead of unital associative $*$--algebras. We present the physical 
motivation for our approach and sketch its mathematical strategy. Moreover, we 
comment on the application of our method to the standard model and the flipped 
$\mathrm{SU(5)} \times \mathrm{U(1)}$--grand unification model. 
\end{abstract}
\vfill

\clearpage 

\section{Physical Motivation}
\label{physmot}

We would like to construct (the classical action of) gauge field theories 
on a space--time manifold $X$ with trivial topology out of the following input 
data:
\vs{-\topsep}
\begin{enumerate}
\item
A unitary matrix Lie group $G$ and its associated gauge group 
$\mathcal{G}=\CX \ot G\,.$ Here, $\CX$ denotes the 
algebra of real--valued smooth functions on $X$.
\label{sybr1}
\vs{-\itemsep} \vs{-\parsep}

\item 
Chiral fermions $\bsj$ transforming under a representation $\t{\pi}_0$ 
of $G\,.$ The induced representation of the gauge group $\mathcal{G}$ is 
$\t{\pi}=\id \ot \t{\pi}_0\,.$ 
\vs{-\itemsep} \vs{-\parsep}

\item
The fermionic mass matrix $\widetilde{\M}\,,$ i.e.\ fermion masses plus 
generalized Kobayashi--Maskawa matrices. 
\label{sybr3}
\vs{-\itemsep} \vs{-\parsep}

\item
Possibly the spontaneous symmetry breaking pattern of $G\,.$  
\label{sybr4}
\end{enumerate}
\vs{-\topsep}
Let us comment on these data. It is common sense that the free Dirac 
action for fermions, 
\eq{
S_F^\mathrm{free}= \int_X dx\; \bsj^* (\sfD + \widetilde{\M}) \bsj~,
   }
is not gauge invariant. In this equation, $\sfD$ is the free Dirac operator 
and $dx$ the volume form on $X\,.$ First, the kinetic term $\bsj^* \sfD \bsj$ 
of the Dirac Lagrangian is not gauge invariant, because $\t{\pi}(\mathcal{G})$ 
does not commute with $\sfD\,.$ Usually, one restores gauge invariance by 
adding gauge fields $\mathrm{A}$ minimally coupled to the fermions. The gauge 
field $\mathrm{A}$ and its action on $\bsj$ are determined by the condition 
that there exist transformations of $\mathrm{A}$ under $\mathcal{G}$ that 
compensate the disturbing part of the transformation of $\bsj^* \sfD \bsj\,.$ 
Second, if the action of only a subgroup $\mathcal{G}_0$ of $\mathcal{G}$ 
commutes with $\widetilde{\M}\,,$ then the mass term 
$\bsj^* \widetilde{\M} \bsj$ of the Dirac Lagrangian is not gauge invariant. 
In this case, one restores gauge invariance by extending the fermionic mass 
matrix to Higgs fields $\widetilde{\M} {+} \Phi$ with appropriate 
transformation behavior. Thus, the gauge invariant fermionic action can be 
written symbolically (i.e.\ up to signs and constants of the order one) as
\eq{
S_F^\mathrm{inv}= \int_X dx\; \bsj^* (\sfD + \widetilde{\M} 
+ \mathrm{A} + \Phi) \bsj~.~~ \label{sin}
   }
Moreover, one wishes to have a dynamics for the fields $\mathrm{A}$ and 
$\Phi\,.$ This is achieved by adding the free bosonic action
\eq{
S_B^\mathrm{free} = \int_X \!\! dx\; ( \langle \bfd \mathrm{A}, \bfd \mathrm{A} 
\rangle_2 + \langle \bfd (\Phi {+} \widetilde{\M}), 
\bfd (\Phi {+} \widetilde{\M}) \rangle_1)~,~~ \label{bpla}
   }
where $\langle ~,~ \rangle_2$ and $\langle ~,~ \rangle_1$ are appropriate 
scalar products. However, the action $S_B^\mathrm{free}$ is not gauge 
invariant, one has to add interaction terms for $\mathrm{A}$ and $\Phi\,.$ 
Moreover, the vacuum expectation value of $\Phi {+} \widetilde{\M}$ must be 
just the mass matrix $\widetilde{\M}$ in order to reproduce the correct 
fermionic sector. This is achieved by adding quartic interaction terms 
$V(\Phi{+}\widetilde{\M})$ such that $\Phi{+}\widetilde{\M}=\widetilde{\M}$ 
is a local minimum of $V(\Phi{+}\widetilde{\M})\,.$ Here, one has to implement 
the desired spontaneous symmetry breaking scheme \ref{sybr4}), which in some 
gauge theories is already determined by the fermionic mass matrix 
$\widetilde{\M}\,.$ However, in extended theories, one may need supplementary 
information on the spontaneous symmetry breaking scheme that is not contained 
in $\widetilde{\M}\,.$ In summary, the invariant bosonic action has the 
symbolic form 
\eqa{rl}{
S_B^\mathrm{inv} = \int_X \!\! dx\; \big( & \langle \bfd \mathrm{A} +
\mathrm{A}^2, \bfd \mathrm{A} + \mathrm{A}^2 \rangle_2 \\[-0.5ex] 
& + \langle (\bfd + \mathrm{A})(\Phi {+} \widetilde{\M}), 
(\bfd + \mathrm{A}) (\Phi {+} \widetilde{\M}) \rangle_1 
+ V(\Phi {+} \widetilde{\M}) \big)\;. \yn \label{bin}
   }
We see that our input data \ref{sybr1}) -- \ref{sybr4}) should 
suffice to reconstruct a complete classical gauge field theory. In particular, 
the fermionic sector determines candidates for the bosonic configuration space. 
Of course, the actions \rf[sin] and \rf[bin] are not unique, but we can fix 
much of the ambiguity by a minimal choice of $\mathrm{A}$ and $\Phi\,.$ 

Usually, the above construction scheme is carried out more or less by hand. 
This is not difficult, for example, in the case of the standard model. However, 
in grand unified theories with very large Higgs multiplets this is a highly 
non--trivial puzzle. One may wish to have a machinery at disposal which is 
able to do this work. This machinery should consist of an algorithm which has 
to be fed with the data \ref{sybr1}) -- \ref{sybr4}) as the input and 
which returns the desired action, in particular, the Higgs multiplets and 
the Higgs potential. This paper is a sketch of such a machinery, which even 
does much more: It also returns tree--level predictions for the masses of 
Yang--Mills and Higgs fields. 

An idea how to find this machinery is inspired by the following observation
\cite{mgv}: The gauge field $\mathrm{A}$ is a vector field and the Higgs field 
$\Phi$ a scalar field. From that point of view, both are completely different 
objects. However, in the above sketch they play precisely the same r\^ole. Both 
$\mathrm{A}$ and $\Phi$ occur via minimal coupling in the fermionic action 
\rf[sin] and restore in this way the gauge invariance. Both have the same type 
of kinetic Lagrangians \rf[bpla]. Both occur as fourth order polynomials in the 
bosonic action \rf[bin]. Moreover, also $\sfD$ and $\widetilde{\M}$ play the 
same r\^ole. All that may be an accident. But accidents have often 
inspired new theories. It might be promissing to search for a new type of 
mathematics that deals with vector and scalar fields in the same way. Such 
mathematics does already exist in form of Alain Connes' non--commutative 
geometry \cite{ac}!

\section{Non--Commutative Geometry}

\subsubsection{General Remarks}

The evolution of non--commutative topology started with Gel'fands discovery 
that the unital $C^*$--algebra $C(X)$ of continuous functions over a compact 
manifold $X$ contains all information about that manifold: Given $C(X)$ one can 
reconstruct the manifold $X$ (up to homeomorphisms) as the set of characters. 
In the other direction, each commutative unital $C^*$--algebra is isomorphic to 
$C(X)$ for a certain compact manifold $X\,.$ This language was transcribed to 
the case that the $C^*$--algebra is not commutative, and one considers general 
$C^*$--algebras as function algebras over ``non--commutative manifolds''. 
This programme, to dualize geometric or topological objects and to deform 
them within the dual picture, has been very successful. It led for instance to 
algebraic K--theory and quantum groups. 

\subsubsection{The Connes--Lott Prescription}

Gel'fands theorem establishes the duality between the function algebra $C(X)$ 
and the topology of $X\,.$ The discovery of Connes \cite{cl,ac} was that, 
taking in addition the Dirac operator acting on the spinor Hilbert space, one 
can also recover the metric properties of $X\,.$ It is possible to reconstruct 
the distance between two points and the de Rham complex. Formalizing this 
method, Connes introduced the basic object of non--commutative geometry, the 
K--cycle or\footnote{We prefer the ancient notation `K--cycle'.} spectral 
triple:
\begin{dfn}
A K--cycle $(\mathcal{A},h,D,\pi,\Gamma)$ over a unital associative 
$*$--algebra $\mathcal{A}$ is given by 
\\
$\mathrm{i)}$ \hfill \parbox[t]{0.95\textwidth}{an involutive representation 
$\pi$ of $\mathcal{A}$ in the algebra $\B(h)$ of bounded operators on a Hilbert 
space $h\,,$ }
\\
$\mathrm{ii)}$ \hfill \parbox[t]{0.95\textwidth}{a (possibly unbounded) 
selfadjoint operator $D$ on $h$ such that $(\one_{\B(h)}+D^2)^{-1}$ is 
compact and for all $a \in \mathcal{A}$ there is $[D,\pi(a)] \in \B(h)\,.$ }
\\[0.5ex]
The K--cycle is called even iff in addition there is a selfadjoint operator 
$\Gamma$ on $h\,,$ fulfilling $\Gamma^2 = \one_{\B(h)}\,,$ 
$\Gamma D + D \Gamma=0$ and $\Gamma \pi(a)-\pi(a)\Gamma=0\,,$ for all 
$a \in \mathcal{A}\,.$	\label{kc} 
\end{dfn} 

Non--commutative geometry (NCG) as sketched above seems to be perfectly adapted 
to the setting \ref{sybr1}) -- \ref{sybr4}): For technical reasons one first 
has to pass from the space--time manifold to a compact Euclidian spin 
manifold $X$. Then, the fermions $\bsj$ constitute the Hilbert space $h\,.$ 
Next, one chooses the selfadjoint operator $D$ of Definition~\ref{kc} to be 
equal to $\sfD+\widetilde{\M}$ on physical fermions $\bsj\,.$ A matrix algebra 
$\mathcal{A}_M$ is chosen in such a way that the gauge group $\mathcal{G}=\CX 
\ot G$ is isomorphic to the group of unitary elements of the algebra 
$\mathcal{A}=\CX \ot \mathcal{A}_M\,.$ The action $\pi=\id \ot \pi_0$ of 
$\mathcal{A}=\CX \ot \mathcal{A}_M$ on $h$ is the extension\footnote{Provided 
that this is possible!} of the group representation $\t{\pi}=\id \ot \t{\pi}_0$ 
of $\mathcal{G}=\CX \ot G$ on the fermions $\bsj\,.$ 
At the very end, one returns to an indefinite metric by a Wick rotation. 
Chiral fermions are obtained by means of a chirality condition via the operator 
$\Gamma\,.$ 

To any K--cycle $(\mathcal{A},h,D,\pi,\Gamma)$ there is canonically associated 
a differential algebra $\Omega^*_D \mathcal{A}\,:$ One considers the universal 
graded differential algebra $\Omega^*\mathcal{A}$ over the algebra 
$\mathcal{A}$ of the K--cycle, 
\al{
\Omega^*\mathcal{A} &= \bigoplus_{n=0}^\infty \Omega^n\mathcal{A}~, &
\Omega^n\mathcal{A} &= \big\{ \tsum_{\alpha} a_{\alpha}^0 \,d a_{\alpha}^1 \, 
d a_{\alpha}^2 \dots d a_{\alpha}^n \big\}~,
      }
where $d$ is the universal differential and $a_{\alpha}^i \in \mathcal{A}\,.$ 
In particular, $\Omega^0\mathcal{A} \cong \mathcal{A}\,.$ One defines a linear 
representation $\pi$ of $\Omega^*\mathcal{A}$ on the Hilbert space $h$ by 
\cite{vg}
\eq{
\pi(a_0 \,d a_1 \,d a_2 \dots d a_n) := \pi(a_0) \cdot [-\iu D, \pi(a_1)] \cdot 
[-\iu D, \pi(a_2)] \cdots [-\iu D, \pi(a_n)]~.
   } 
One remarks that $\pi(\Omega^*\mathcal{A})$ is not a differential 
algebra. Fortunately, this defect can be repaired, and the canonical graded 
differential algebra is 
\eqa{rclrcl}{
\Omega^n_D \mathcal{A} &=& \bigoplus_{n=0}^\infty \Omega^n_D\mathcal{A}~, 
\qquad{} & 
\Omega^n_D\mathcal{A} &:=& \Omega^n\mathcal{A}\,/\, ((\ker \pi + d \ker \pi) 
\cap \Omega^n\mathcal{A} ) \npb \\[-2ex] &&& 
&\cong& \pi(\Omega^n\mathcal{A})\,/\, \pi( d \ker \pi 
\cap \Omega^n\mathcal{A} ) ~. \yn
   }
For the physically interesting case of even K--cycles over a subalgebra of 
$\CX_\C \ot \mat{F}$ and generalized Dirac 
operators of the form $D=\sfD \ot \one_F + \g[5] \ot \M\,,$ a generally 
applicable construction of $\Omega^*_D \mathcal{A}$ has been given in 
\cite{kppw}.
The non--commutative gauge potential is an element of $\Omega^1_D\mathcal{A}$ 
and the field strength an element of $\Omega^2_D\mathcal{A}\,.$ Using invariant 
scalar products one defines bosonic and fermionic actions \cite{ac,cl}. 
A further improvement is a new spectral action principle 
\cite{cc1,cc2} that gives a coupling of the Yang--Mills 
(--Higgs) action to Einstein plus Weyl gravity.

\subsubsection{Application to the Standard Model}

This NCG--prescription has proved very successful in reformulating the standard 
model. There exists an ``old scheme'' initiated by Connes and Lott in 
\cite{cl}, see also \cite{ac, ks4, ks, vg, iks1}, 
and a ``new scheme'' based upon real structures introduced by Connes in 
\cite{acr}, see also \cite{iks, mgv, cis} 
for the application to model building. The algebra $\mathcal{A}$ and its group 
of unitary elements $\mathcal{U}(\mathcal{A})$ are given by
\eqas{rcl}{
\mathcal{A}_\mathrm{old} &=& \CX \ot \big((\mathbbm{H} \op \C) \op 
(\mat{3} \op \C) \big)~,~~ \\
\mathcal{U} (\mathcal{A}_\mathrm{old} ) &=& \CX \ot (\SU2 \times \U1 \times \U3 
\times \U1) ~, \\
\mathcal{A}_\mathrm{new} &=& \CX \ot (\mathbbm{H} \op \C \op \mat{3}) ~,~~ \\
\mathcal{U} (\mathcal{A}_\mathrm{new} ) &=& 
\CX \ot (\SU2 \times \U1 \times \U3) ~.
	}
The additional $\U1$--groups are eliminated by unimodularity conditions. 
The most important improvement compared with the usual formulation of the 
standard model is that the non--commutative gauge potential contains both the 
$\su3 \op \su2 \op \u1$ Yang--Mills fields and the complex Higgs doublet. 
Moreover, the bosonic action contains the Yang--Mills Lagrangian, the covariant 
derivatives of the Higgs fields and the Higgs potential in a unified form. The 
fermionic action unifies the gauge field couplings with the Yukawa--couplings. 
Numerically, one gets a very promissing ``fuzzy'' relation between the mass of 
the $W$ boson and the mass of the top quark, and the prediction for the mass of 
the Higgs field is compatible with LEP precision experiments, see 
\cite{ac,acr,iks,iks1,cis,mgv,ks4}.

\subsubsection{The Mainz--Marseille Model}

There exists a different NCG--formulation of the standard 
model \cite{cev,ces,chps,hps} elaborated by groups in Mainz and Marseille. This 
formulation leads to the same unification of the Yang--Mills and Higgs sectors 
in the bosonic and fermionic actions. The essential mathematical difference is 
the use of the graded Lie algebra $\Lambda^* \ot \su{2|1}$ of differential 
form--valued matrices as the starting point instead of K--cycles and 
differential algebras constructed thereof in the Connes--Lott prescription. 
The essential physical difference is that the purely bosonic sector of the 
standard model can be formulated. This is in contrast to the Connes--Lott 
model, where the bosonic sector can only be reproduced if at least two 
generations of fermions occur in nature (which is the case, of course). The 
Mainz--Marseille model yields no relations between fermion and boson masses, 
but an interesting relation between the Cabibbo angle and quark masses can be 
obtained \cite{ces, hs}. 

The inseparable tie between bosons and fermions in the Connes--Lott model, 
which is responsible for relations between fermion and boson masses obtained 
in that model, has been criticized by the Mainz--Marseille group, mainly for 
two reasons: First, purely bosonic theories are mathematically interesting as 
well. Second, relations between fermion and boson masses do not survive the 
usual quantization procedure. However, there exist examples where parameter 
relations that are not stemming from a symmetry of the theory are respected on 
quantum level, see \cite{z}. Thus, our point of view is to consider the 
interpretation of the mass relations in the Connes--Lott model as a challenge 
for the future. 

\subsubsection{Non--Commutative Geometry and Grand Unification}

The overwhelming success of non--commutative geometry leads to the expectation 
that its application to other gauge field theories should be not difficult. 
However, if one follows the Connes--Lott prescription one runs into certain 
problems. It was shown in \cite{lmms} that, besides the standard model, there 
are only two more or less realistic models which can be constructed within the 
above understanding of NCG: the $\SU4_{PS} \times \SU2_L \times \SU2_R$--model 
and the $\SU3_C \times \SU2_L \times \SU2_R \times \U1_{B-L}$--model. However, 
if one additionally demands a real structure \cite{acr} for the K--cycle, then 
also these two models are ruled out. The only more or less realistic physical 
model that is compatible with the most elegant NCG--prescription is the 
standard model! It is certainly to early to judge from 
experimental results whether the standard model is correct or not. At least 
there exist good reasons \cite{l} why one could be interested in Grand Unified 
Theories (GUT's): GUT's explain the quantization of electric charge, yield a 
fairly well prediction for the Weinberg angle, explain the convergence of 
running coupling constants at high energies, include massive neutrinos to 
solve the solar neutrino problem, produce the observed baryon asymmetry of the 
universe, etc. Unfortunately, the results of \cite{lmms} imply that one needs 
additional structures or different methods for a NCG--formulation of these 
models.

The perhaps most successful NCG--approach towards grand unification was 
proposed by Chamseddine, Felder and Fr\"ohlich. In the $\SU5$--model 
\cite{cff1, cff2}, the authors start to construct an auxiliary K--cycle. Within 
this framework they construct the bosonic sector. Then they interpret some of 
these bosonic quantities as Lie algebra valued and consider Lie algebra 
representations on the physical Hilbert space to obtain the fermionic sector. 
This procedure is a systematic realization of the gauge theory construction 
programme set up at the beginning. However, an aesthetic shortcoming of that 
approach is the auxiliary character of the K--cycle, which of course is 
inevitable in view of \cite{lmms}. The $\mathrm{SO(10)}$--model \cite{cf} by 
Chamseddine and Fr\"ohlich fits well\footnote{Nevertheless, 
the use of Lie algebras instead of algebras could probably justify certain 
assumptions made in \cite{cf}.} into the NCG--scheme. The reason 
why this model was excluded in \cite{lmms} is that only models possessing 
complex fundamental irreducible representations were admitted in that article. 

It turns out that only a slight modification of the Connes--Lott prescription 
enables the formulation of a large class of physical models without additional 
structures. A sketch of that formulation and of its application to 
interesting physical models is the concern of this paper.

\section{Non--Associative Geometry}

Let us investigate why the most elegant NCG--prescription is so restrictive to 
admissible models. The obstruction is the extension of the representations of 
the gauge group $\mathcal{G}=\CX \ot G$ to representations of the unital 
associative $*$--algebra $\mathcal{A}=\CX \ot \mathcal{A}_M$ containing 
$\mathcal{G}$ as the set of unitary elements. That $\t{\pi}=\id \ot \t{\pi}_0$ 
is a representation of $\mathcal{G}$ on the Hilbert space $h$ means that 
\eq{
\t{\pi}_0 (g_1) \, \t{\pi}_0 (g_2) = \t{\pi}_0 (g_1 g_2) ~, \quad
\forall g_1,g_2 \in G~.
   }
The representation $\t{\pi}_0$ of the matrix group $G$ should coincide with the 
representation $\pi_0$ of the matrix algebra $\mathcal{A}_M$ on the subset 
$G \subset \mathcal{A}_M\,,$
\eq{
\pi_0 (g_1) \, \pi_0 (g_2) = \pi_0 (g_1 g_2) ~, \quad \forall g_1,g_2 \in G 
\subset \mathcal{A}_M~.
\label{pgpg}
   }
It is perhaps not the problem to extend the multiplication rule \rf[pgpg] 
to the entire matrix algebra $\mathcal{A}_M\,.$ The essential problem is that 
this extension must be compatible with linear operations, 
\eq{
\lambda_1 \pi_0 (a_1) + \lambda_2 \pi_0 (a_2) 
= \pi_0 (\lambda_1 a_1 + \lambda_2 a_2) ~, \quad
\forall a_1,a_2 \in \mathcal{A}_M ~,~~ \forall \lambda_1,\lambda_2 \in \R ~.
\label{addp}
   }
Addition and multiplication by scalars are not defined on $G\,,$ and the 
representation $\t{\pi}_0$ does not care whether it is linear or not. A priory, 
there are two types of irreducible representations that fulfil \rf[addp]: the 
identity and~-- in the case of real algebras~-- the complex conjugation. In 
general, this is all what is possible. We see: The reason why the most elegant 
NCG--prescription \cite{acr} is so restrictive is that it is compatible only 
with linear representations of the matrix group. Most of the grand unified 
theories are not of that type. 

Fortunately, our observation also shows the way how to overcome the 
restriction: We propose to linearize the matrix group, which means to work 
within the tangent space at a fixed group element, for instance the unit 
element. The tangent space at the unit element is isomorphic to the Lie algebra 
$\f[a]$ of $G\,.$ Thus, \emph{the Lie algebra $\f[g]=\CX \ot \f[a]$ of the 
gauge group $\mathcal{G}=\CX \ot G$ is the correct object to use, not an 
algebra extending $\mathcal{G}\,.$} The linearized group multiplication is 
described by the commutator of Lie algebra elements. It is clear that the 
representation of a Lie group induces a representation of its Lie algebra. The 
point is that this Lie algebra representation is always linear. 

In analogy to the procedure in non--commutative geometry we formalize our 
observation. We simply replace in Definition~\ref{kc} the unital associative 
$*$--algebra $\mathcal{A}$ by a skew--adjoint Lie algebra $\f\,.$ The outcome 
can no longer be called a K--cycle; I propose the name ``L--cycle'', where 
the letter L stands for Lie (and it is the next letter in the alphabet). We 
also cannot keep the name non--commutative geometry, because a Lie bracket is 
always (anti--)commutative. I suggest the name ``non--associative geometry'', 
because~-- in general~-- the Lie bracket is not associative. However, it must 
be stressed that our approach can not be applied to general non--associative 
algebras. Thus, the title could be misleading, but any title carries the risk 
of wrong associations. 
 
The point of departure in our approach is the following definition: 
\begin{dfn} 
An L--cycle $(\f,h,D,\pi,\Gamma)$ over a skew--adjoint Lie algebra $\f$ is 
given by 
\\
$\mathrm{i)}$ \hfill \parbox[t]{0.95\textwidth}{an involutive representation 
$\pi$ of $\f$ in the Lie algebra $\B(h)$ of bounded operators on a Hilbert 
space $h\,,$ i.e. $(\pi(a))^*=\pi(a^*) \equiv -\pi(a) \,,$ for any 
$a \in \f\,,$ } 
\\
$\mathrm{ii)}$ \hfill \parbox[t]{0.95\textwidth}{a (possibly unbounded) 
selfadjoint operator $D$ on $h$ such that $(\id_h{+}D^2)^{-1}$ is compact and 
for all $a \in \f$ there is $[D,\pi(a)] \in \B(h)\,,$ where $\id_h$ denotes 
the identity on $h\,.$} 
\\
$\mathrm{iii)}$ \hfill \parbox[t]{0.95\textwidth}{a selfadjoint operator 
$\Gamma$ on $h\,,$ fulfilling $\Gamma^2 = \id_h\,,$ $\Gamma D {+} D \Gamma=0$ 
and $\Gamma \pi(a) {-} \pi(a)\Gamma=0\,,$ for all $a \in \f\,.$ } 
\label{lcc} 
\end{dfn} 

It seems obvious that the concept of non--associative geometry 
is perfectly adapted to the setting \ref{sybr1}) -- \ref{sybr4}) at the 
beginning\footnote{There can occur obstructions and modifications if 
Abelian Lie groups are present. In particular, a purely Abelian gauge field 
theory can be constructed only with partial success.  
In some cases, Abelian Lie algebras are automatically generated. If such a Lie 
algebra is desired, one can omit this part when deriving the Lie algebra 
$\f[g]$ out of $\mathcal{G}\,.$ \label{fnu1}}:
As in non--commutative geometry we start with the construction of the Euclidian 
gauge field theory. Again, the Euclidian fermions $\bsj$ constitute our 
Hilbert space $h\,.$ For technical reasons it may sometimes be necessary to 
work with several copies of the fermions. The Lie algebra $\f[g]=\CX \ot \f[a]$ 
is simply the Lie algebra of the gauge group $\mathcal{G}\,,$ 
up to possible modifications if $\U1$--groups occur (see footnote~\ref{fnu1}).
We assume that $X$ has a trivial topology in order to avoid discussions of 
transition functions between different charts of the manifold. 
The Lie algebra representation $\pi=\id \ot \h{\pi}$ is just the differential 
$\t{\pi}_*$ of the group representation $\t{\pi}=\id \ot \t{\pi}_0\,.$ The 
selfadjoint operator $D$ is chosen in such a way that on physical fermions it 
equals $\sfD + \widetilde{\M}\,.$ The operator $\Gamma$ represents the 
chirality properties of the fermions. Finally, one returns to Minkowski space 
by a Wick rotation and imposes a chirality condition for the fermions $\bsj$ 
by means of $\Gamma.$ 

The programme of non--associative geometry is clear: 
We ``simply'' have to transcribe the Connes--Lott prescription of 
non--commutative geometry to our case. However, this is not as easy as one 
probably expects. The associativity of the algebra and the existence of a unit 
element are very powerful tools. Without them we are forced to go long detours 
where non--commutative geometry uses short cuts.

\section{The General Scheme}

Now for the sketch of the construction in the general context, without relation 
to physical models. A detailed exposition of our techniques can be found in 
\cite{rw2}. In analogy to the first step in 
non--commutative geometry we enlarge our Lie algebra $\f[g]$ to a universal 
graded differential Lie algebra $\W{*}\,.$ One can imagine $\W{*}$ as the 
set of repeated graded commutators of $\f[g]$ and $d \f[g]\,,$ where 
$d\f[g]$ is a second copy of $\f[g]\,.$ Thus, elements $\omega \in \W{*}$ have 
the form 
\eq{
\omega=\sum_{\alpha,z \geq 0} [v^z_{\alpha} ,[v^{z-1}_{\alpha} ,[ \dots , 
[v^1_{\alpha}, v^0_{\alpha} ] \dots ]]]~, \quad \mbox{finite sum}~,
   }
where $v^i_{\alpha}$ either belongs to $\f[g]$ or $d\f[g]\,.$ The vector space 
$\W{*}$ is $\N$--graded. The homogeneous element 
$[v^z ,[v^{z-1},[ \dots ,[v^1, v^0 ] \dots ]]]$ belongs to $\W{n}$ 
iff $n$ elements of $\{v^0,\dots,v^z\}$ belong to $d\f[g]\,.$ The graded 
commutator $[~,~]$ is compatible with that grading structure; one has 
$[\W{k},\W{l}] \subset \W{k+l}\,.$ Moreover, $[~,~]$ respects the usual 
graded antisymmetry and the graded Jacobi identity. The symbol $d$ is extended 
to a graded differential on $\W{*}\,,$ it is nilpotent and obeys the graded 
Leibniz rule. The graded Lie algebra $\W{*}$ is universal in the following 
sense: Each graded differential Lie algebra generated by $\pi(\f[g])$ and 
$\mathrm{d} \pi(\f[g])$ can be obtained by factorization of $\W{*}$ with 
respect to a differential ideal. For instance, the information contained in 
an L--cycle determines uniquely such a differential ideal. Thus, there is a 
canonical graded differential Lie algebra $\D{*}$ associated to an L--cycle. 

To find this differential Lie algebra, we represent $\W{*}$ on the Hilbert 
space $h\,,$ using the data specified in the L--cycle. This representation 
extends the representation $\pi$ of the L--cycle and is defined by 
\eqas{rcl}{
\pi(da) &=& [-\iu D,\pi(a)]~,~~ \\
\pi([\omega^k,\t{\omega}^l]) &=& [\pi(\omega^k),\pi(\t{\omega}^l)]_g :=
\pi(\omega^k) \pi(\t{\omega}^l) -(-1)^{kl} \pi(\t{\omega}^l) \pi(\omega^k) ~,
       }
for $a \in \f[g]\,,$ $\omega^k \in \W{k}$ and $\t{\omega}^l \in \W{l}\,.$ Here, 
it is essential to have the grading operator $\Gamma\,,$ which detects the 
correct sign for $(-1)^{kl}\,.$ 

As one expects from non--commutative geometry, the representation $\pi$ does 
not transport the differential $d$ on $\W{*}$ to a differential on $\P{*}\,.$ 
To cure this, we use the same trick as in non--commutative geometry. One shows 
that 
\eq{
\mathcal{J}^*\! \f[g] = \ker \pi + d \ker \pi  \subset \W{*}
   }
is a graded differential ideal of $\W{*}\,.$ Factorizing out the ``junk'' 
$\mathcal{J}^*\! \f[g]$ we obtain the graded differential Lie algebra 
$\D{*}\,,$
\al{
\D{*} &=\bigoplus_{n=0}^\infty \D{n}~,& \D{n} &
=\dfrac{\W{n}}{\mathcal{J}^n \!\f[g]} \cong 
\dfrac{\P{n}}{\pi(\mathcal{J}^n \!\f[g])} ~.~~
   }
The differential and the commutator are defined as usual for equivalence 
classes. 

It is extremely useful to introduce a linear map $\sigma$ from $\W{*}$ 
to (possibly unbounded) operators on $h\,.$ The operator $\sigma$ is odd with 
respect to the $\Z_2$--grading and is within the same notations as before 
defined by 
\eqas[sww]{rcl}{
\sigma(a) &=& 0~, \qquad \sigma(da)=[D^2,\pi(a)]~,~~ \npb \\
\sigma([\omega^k,\t{\omega}^l]) &=& [\sigma(\omega^k),\pi(\t{\omega}^l)]_g 
+ (-1)^k [\pi(\omega^k), \sigma(\t{\omega}^l) ]_g ~.~~	
	}
The importance of the map $\sigma$ is that it measures the defect if one 
represents the universal differential $d$ by graded commutators with 
$- \iu D\,,$ 
\eq{
\pi(d \omega^k) = [-\iu D,\pi(\omega^k)]_g + \sigma(\omega^k)~, \quad 
\omega^k \in \W{k}~. \label{pdw}
   }
In particular, taking $\omega^k \in \ker \pi\,,$ we get
\eq{
\J{k+1}=\{~ \sigma(\omega^k)~,~~ \omega^k \in \W{k} \cap \ker \pi~\}~.
   }
This characterization of $\J{*}$ is especially convenient, because 
$\sigma(\omega^k)$ is derived successively from lower degrees, see \rf[sww]. 
Indeed, this is the way how we can eventually compute $\J{*}$: The real 
problem is to find $\sigma(\W{1})\,.$ Then we derive for $k \geq 2$ by 
induction a formula for $\sigma(\omega^k)$ for given $\pi(\omega^k)\,.$ 
Clearly, $\sigma(\omega^k)$ is not uniquely defined by $\pi(\omega^k)\,,$ and 
this ambiguity is nothing but $\J{k+1}\,.$ However, the explicit realization of 
this line is not done within a couple of pages. We also point out that, once 
knowing $\sigma(\omega^k)\,,$ formula \rf[pdw] provides the explicit 
differentiation rule for elements of $\D{*}\,.$ 

In non--commutative geometry, all work is done at this point. There, the 
connection form is simply an element of $\Omega^1_D \mathcal{A}$ and the 
curvature an element of $\Omega^2_D \mathcal{A}\,.$ It is straightforward to 
write down the fermionic and bosonic actions. In non--associative geometry, the 
situation is different. If one tries to find a reasonable definition for the 
connection (the covariant derivative), one encounters more freedom than one 
expects. Moreover, it is not possible to describe gauge field theories 
containing $\U1$--groups if one takes $\D{1}$--valued connection forms. 
Therefore, an additional structure is necessary: Not the graded differential 
Lie algebra $\D{*}$ is the correct space where the connection form and the 
curvature live, but the space of certain graded Lie endomorphisms of 
$\D{*}\,.$ This is not completely unreasonable. For instance, connections 
within the framework of finite projective modules \cite{mrw2} are of a similar 
type. Formally, we introduce the space $\H{*}= \bigoplus_{n \in \N} \H{n}$ of 
certain graded Lie homomorphisms of $\P{*}\,.$ The space $\H{n}$ consists of 
linear (possibly unbounded) operators on $h$ of $\Z_2$--degree $n \mod 2\,,$ 
which raise the $\N$--degree of $\P{*}$ and $\J{*}$ by $n\,,$
\al{
[\H{n},\P{k}]_g &\subset \P{k+n}~, & [\H{n},\J{k}]_g &\subset \J{k+n}~.~~  
\label{hp}
  }
Factorizing $\H{*}$ with respect to its graded centre $\tcc{*}$ in $\P{*}$ and 
the ideal $\J{*}\,,$ we obtain the graded Lie algebra 
\al{
\hH{*} &:= \bigoplus_{n \in \N} \hH{n} ~,& 
\hH{n} &:= \H{n}\,/\, (\J{n} + \tcc{n})~.
   }
The differential and the commutator on 
$\hH{*}$ are defined as usual for dual spaces: via the graded Leibniz rule and 
the graded Jacobi identity. From our definitions it is clear that
\al{
\P{n} &\subset \H{n}~, & \D{n} &\subset \hH{n}~.~~
\label{cpl}
   }
In some sense, this framework is an extension of the primary spaces $\P{*}$ and 
$\D{*}\,.$ 

The formal definition of a connection on L--cycles is given in \cite{rw2}. 
Here, we shall only quote the result: A connection $\nabla$ acting on $\D{*}$ 
is closely related to the covariant derivative $\nabla_h$ acting on the 
Hilbert space $h\,.$ The general form of these two objects is 
\al{
\nabla_h &=-\iu D + \rho~,& \nabla &=d + [\t{\rho}\,,~.~]_g~,& 
\rho &\in \H{1}~, ~~ \t{\rho} := \rho + \tcc{1} \in \hH{1}~.
   }
The Lie homomorphism $\rho$ is called the connection form (gauge potential). 
The curvature (field strength) of the connection $\nabla$ is 
\al{
\nabla^2 &=[\theta,~.~]~,&  \theta &=d \t{\rho} 
+ \tfrac{1}{2} \{\t{\rho},\t{\rho}\} \in \hH{2}~.~~
   }
We see that our formulae look very similar to what one knows from NCG or 
classical gauge field theory. However, we have no control over the space of 
connections in that general context. All what we know is that elements of 
$\D{1}$ are possible connection forms, but it is completely unclear what else. 
Also the operations $d\t{\rho}$ and $\{\t{\rho},\t{\rho}\}$ are difficult to 
perform, because they are only indirectly defined. It is a visible 
complication compared with non--commutative geometry to find not only 
$\D{*}$ but also $\hH{*}$ (up to second degree).

The group $\mathcal{U}(\f[g])$ obtained via the exponential mapping of a 
neighbourhood of the zero element of $\H{0}$ plays the r\^ole of a gauge group 
in our approach. Comparing for a physical model this group with the original 
gauge group $\mathcal{G}$ we had started with, we see that the 
global topology of $\mathcal{G}$ cannot always be reconstructed. But for most 
physical applications it suffices to know the gauge group locally. 
One can define an adjoint representation $\mathrm{Ad}$ of $\mathcal{U}(\f[g])$ 
on $\D{*}\,.$ Local gauge transformations are given by
\eqas[nadu]{rclrclrcl}{
\nabla &\mapsto&  \Ad{u} \nabla \Ad{u^*}~,\qquad{} &
\nabla_h &\mapsto&  u \nabla_h u^* ~,\qquad{} \\
\rho &\mapsto& u d u^{-1} + u \rho u^* ~,\qquad{} & 
\theta &\mapsto&  \Ad{u} (\theta) ~, \qquad{} & \bsj &\mapsto&	u \bsj ~,~~ 
   }
where $u \in \mathcal{U}(\f[g])$ and $\bsj \in h\,.$ The bosonic and fermionic 
actions are defined in the same way as in non--commutative geometry: 
Using the Dixmier trace $\Tr_{\omega}$ we define the bosonic action
\eq{
S_B(\nabla) := \min_{j^2 \in \tcc{2} + \J{2}} 
\Tr_{\omega} ((\theta_0 + j^2)^2\, |D|^{-\mathrm{d}}) ~,  \label{sbn}
   }
where $\theta_0 \in \H{2}$ is any representative of $\theta\,.$    
For the fermionic action we use the scalar product on the Hilbert space:
\eq{
S_F(\bsj,\nabla_h) := \langle \bsj ,\iu \nabla_h \bsj \rangle_h ~, \quad
\bsj \in h~.
   }
Both $S_B$ and $S_F$ are invariant under gauge transformations \rf[nadu]. 

\section{Functions $\ot$ Matrices}

In physical applications one is especially interested in the case that the 
Lie algebra $\f[g]$ is the tensor product of the algebra of functions on the 
space--time manifold $X$ and a matrix Lie algebra $\f[a]\,.$ We are able to 
handle this situation. However, it turns out that we must impose restrictions 
on the matrix Lie algebra. If $\f[a]$ is semisimple then there are no problems 
at all. The situation that $\f[a]$ is Abelian can not be satisfactory treated. 
We are able to deal with L--cycles over the Lie algebra 
\eq{
\f[g]=\CX \ot (\f[a]' \op \f[a]'')~,
   }
where $\CX$ is the algebra of real smooth functions over the (four 
dimensional) space--time manifold, $\f[a]'$ is a semisimple Lie algebra and 
$\f[a]''$ an optional Abelian Lie algebra. For $\f[a]''$ we have to impose 
constraints on the representations. Remarkably, for the models I considered so 
far, the $\u1$--representations realized in nature are admissible. The Hilbert 
space is 
\eq{
h=L^2(X,S) \ot \C^F~,
   }
where $L^2(X,S)$ is the Hilbert space of square integrable sections on the 
spinor bundle over $X\,.$ The representation $\pi$ of $\f[g]$ on $h$ is given 
by 
\eq{
\pi=\id \ot \h{\pi}~,
   }
where $\h{\pi}$ is a representation of $\f[a]' \op \f[a]''$ on $\C^F\,.$ The 
selfadjoint operator $D$ of the L--cycle is 
\eq{
D=\sfD \ot \one_F + \g \ot \M~,
   }
where $\sfD$ and $\g$ are the Dirac operator of the spin connection and 
the chirality operator on $L^2(X,S)\,.$ Moreover, $\M$ is a 
symmetrical complex $F \times F$--matrix such that there exists a symmetrical 
$F \times F$--matrix $\h{\Gamma},$ fulfilling $\h{\Gamma}^2=\one_F\,,$ 
$\M \h{\Gamma} = - \h{\Gamma} \M$ and $\h{\pi}(a) \h{\Gamma} 
= \h{\Gamma} \h{\pi}(a) \,,$ for all $a \in \f[a]\,.$ Then, the chirality 
operator is 
\eq{
\Gamma= \g \ot \h{\Gamma}~.
   }
As mentioned before, the representation $\pf[a'']$ is not arbitrary, we have a 
constraint relation between $\M$ and $\pf[a'']\,,$ see \cite{rw2}. Observe that 
the tuple $(\f[a],\C^F,\M,\h{\pi},\h{\Gamma})$ itself forms an L--cycle. In 
some sense, the L--cycle $(\f[g],h,D,\pi,\Gamma)$ is the product of the Dirac 
K--cycle $(\CX,L^2(X,S),\sfD,\g)$ with the matrix L--cycle 
$(\f[a],\C^F,\M,\h{\pi},\h{\Gamma})\,.$ 

One may ask how the spaces $\P{*},\J{*}$ and $\D{*}$ depend on the geometric
objects of the underlying Dirac K--cycle and the matrix L--cycle.
It turns out that $\P{*},\J{*}$ and 
$\D{*}$ can be universally written as a sum of tensor products of 
differential forms of homogeneous degree (partly coboundaries only) with 
certain commutators and anticommutators of homogeneous subspaces of $\p{*}$ and 
$\jj{*}\,.$ Thus, if one has complete knowledge of $\p{*}$ and $\jj{*}\,,$ then 
also $\P{*},\J{*}$ and $\D{*}$ are known. The formulae of lowest degree read:
\eqa{rcl}{
\P{0} &=& \Lambda^0 \ot (\pf[a'] \op \pf[a''])~,~~ \npb \\
\P{1} &=& (\Lambda^1 \ot \pf[a']) \op (B^1 \ot \pf[a'']) 
\op (\Lambda^0 \g \ot \p{1}) ~,~~ \npb \\
\P{2} &=& (\Lambda^2 \ot \pf[a']) \op (\Lambda^1 \g \ot \p{1}) 
\op (\Lambda^0 \ot (\p{2} + \{\pf,\pf\}))~,~~ \\
\J{0} &=& 0~, \qquad{} \J{1}=0~, \qquad{} 
\J{2}=\Lambda^0 \ot (\jj{2} + \{\pf,\pf\}) ~,  \\
\D{0} &=& \P{0}~, \qquad \qquad \D{1}=\P{1}~,~~\npb \yn \\ 
\D{2} &=& (\Lambda^2 \ot \pf[a']) \op (\Lambda^1 \g \ot \p{1})  \npb \\
&& \op (\Lambda^0 \ot \big((\p{2} + \{\pf,\pf\}) \mod (\jj{2} + \{\pf,\pf\}) 
\big)) ~.~~
     }
Here, $\Lambda^k$ is the space of $k$--differential forms, 
$B^1=\bfd \Lambda^0 \subset \Lambda^1$ the space of 1--co\-boun\-daries and 
\eq{
\{\pf,\pf\} = \{ ~ \tsum_{\alpha} \{ \h{\pi}(a_{\alpha}) , 
\h{\pi}(\t{a}_{\alpha}) \}~,~~
a_{\alpha},\t{a}_{\alpha} \in \f[a]~,~~\mbox{finite sum}~ \}~.~~
   }
For higher degrees, the formulae for the matrix part belonging to a fixed space 
of $k$--differential forms become more and more complicated. Corresponding 
formulae in NCG are less difficult, because an associative algebra does not 
care, at which sites in the product $\omega_1 \circ \omega_2 \circ \dots \circ 
\omega_n$ one inserts brackets distinguishing commutators and anticommutators. 
As it can be seen, the Abelian Lie algebra $\f[a]''$ plays a special r\^ole. 
For instance, if the connection form $\rho$ belongs to $\D{1}\,,$ then the 
field strength of a $\u1$--gauge field is always zero. That $\u1$--gauge 
fields can have a non--vanishing field strength in our theory is due to the 
extension of $\D{1}$ to $\hH{1}\,.$ 

An additional feature of L--cycles over functions $\ot$ matrix Lie algebra 
is the possibility to consider local connections. For local connections, the 
connection form $\rho$ commutes with functions. Therefore, it has the 
decomposition
\eq{
\rho \in (\Lambda^1 \ot \bbr{0}) \op (\Lambda^0 \g \ot \bbr{1}) ~,~~ 
\label{rsl}
   }
where $\bbr{0}$ and $\bbr{0}$ are certain subspaces of $\mat{F}\,.$ The 
defining equations \rf[hp], decomposed according to their differential form 
degree, yield certain equations for commutators and anticommutators of 
$\bbr{0}$ and $\bbr{1}$ with $\p{*}$ and $\jj{*}\,.$ These equations and 
$\Z_2$--grading properties and involution identities make it possible to find 
the space of gauge potentials \rf[rsl]. Moreover, one also gets a decomposition 
for the ideal $\cJ{2}:=\tcc{2} + \J{2} $ commuting with functions, which we 
need to write down the bosonic action \rf[sbn]: 
\eq{
\cJ{2} = (\Lambda^0 \ot \cc{2}) \op (\Lambda^1 \g \ot \cc{1}) 
\op (\Lambda^2 \ot (\cc{0} {+} \jj{2} {+} \{\pf,\pf\}))~.
   }
Again, one finds certain equations between $\cc{i}$ and $\p{*}$ that 
make it possible to determine $\cJ{2}\,.$ For the computation of the bosonic 
action one makes use of the fact that in the present situation one can express 
the Dixmier trace by a combination of the usual trace over the matrix 
structures (including gamma matrices) and integration over the space--time 
manifold. 

\section{Electrodynamics and Standard Model}

One can try to formulate the chiral spinor electrodynamics within our approach. 
However, since the Lie algebra to use is purely Abelian, there occur certain 
problems. It is no problem to get the correct fermionic action. In particular, 
the photon has the usual properties and a non--vanishing classical curvature. 
Nevertheless, in our approach we get a vanishing curvature and, therefore, no 
bosonic action.

The reformulation of the standard model \cite{rw3} is more successful. The 
L--cycle is the direct transcription of the physical 
situation. Clearly, the Lie algebra to use is $\CX \ot (\su3 \op \su2 \op \u1) 
\,.$ We can formulate the standard model with or without right neutrinos. 
For a generic mass matrix, the generalized gauge potential $\rho$ contains the 
usual Yang--Mills fields of the standard model and one complex Higgs doublet. 
The bosonic Lagrangian includes the Yang--Mills part, the covariant derivative 
of the Higgs fields and the well--known quartic Higgs potential. Three Higgs 
components are absorbed by the Higgs mechanism and give mass to the $W^\pm$ 
and $Z$ bosons. One massive scalar Higgs field survives. In the same way as in 
non--commutative geometry we obtain tree--level predictions for all bosonic 
masses. For the simplest scalar product we find in the case that right 
neutrinos are included 
\al{
m_W &= \tfrac{1}{2} m_t~,&  m_Z &=m_W/\cos \theta_W~,& 
\sin^2 \theta_W &= \tfrac{3}{8}~,& m_H &=\tfrac{3}{2} m_t~.
   }
Without right neutrinos, the only modification is $m_H =\sqrt{\tfrac{43}{20}} 
m_t\,.$ Here, $m_t,m_W,m_Z,m_H$ are the masses of the top quark, the $W$ 
bosons, the $Z$ boson and the Higgs boson. The photon and the gluons remain 
massless. The Weinberg angle $\theta_W$ coincides with the $\SU5$--GUT 
prediction. Moreover, we get the same coupling constants for the weak and 
strong interactions. In the Connes--Lott formulation of non--commutative 
geometry one uses the algebra $\mathcal{A}= \CX \ot (\mat{3} \op \mathbbm{H} 
\op \C)$ to derive the standard model, together with a rather complicated 
representation of $\mathcal{A}\,.$ For the simplest scalar 
product\footnote{In the meantime one prefers to use the whole class of 
compatible scalar product to obtain ``fuzzy mass relations'', see \cite{cis, 
iks1, iks, mgv}.}, the numerical results are \cite{ks}
\al{
m_W &= \tfrac{1}{2} m_t~,& m_Z &=m_W/\cos \theta_W~,& 
\sin^2 \theta_W &=\tfrac{12}{29}~,& m_H &= \sqrt{\tfrac{69}{28}} m_t 
\approx 1.57 \,m_t~.
   }
Thus, we see that the predictions from non--associative and non--commutative 
geometry do not differ very much. 

\section{The Flipped $\SU5 \times \U1$--Grand Unification Model}

This section is a summary of our analysis \cite{rw4} of the flipped 
$\SU5 \times \U1$--model. For the classical treatment of that model see 
\cite{dkn}. 

\subsubsection{The Matrix L--Cycle}

The matrix L--cycle is given by the following data: The matrix Lie algebra is 
$\f[a]=\su5\,.$ Nevertheless, we will obtain an additional $\u1$--gauge field 
and $\U1$--gauge transformations due to the extension of $\P{1}$ to $\H{1}\,.$ 
Remarkably, the representation of that $\u1$--gauge field on the fermionic 
Hilbert space is unique and realized in nature\footnote{This is a purely 
algebraic result, for which I have no geometric interpretation. I suppose that 
this has something to do with anomaly--freedom of the model.}! The internal 
Hilbert space is $\C^{192}\,.$ This means that we must deal with huge matrices, 
a problem which should not be underestimated. The strange number 
$192= 4 \cdot 48$ arises because there are 48 fermions in nature (including 
right neutrinos), and we need four copies of them: Two copies because we need 
particles and antiparticles in one representation (the $\SU5$ exchanges 
particles and antiparticles -- proton decay!), and an additional doubling 
to include the essential grading operator. The 48 fermions occur in three 
generations, each generation contains 16 fermions. These 16 fermions are 
assigned to the $\su5$--representations $\ul{10},{\ul5}^*,\ul1\,.$ 
Now, for $a \in \su5$ we define the representation $\h{\pi}$ of the Lie algebra 
$\su5$ of our matrix L--cycle in terms of $48 \times 48$--block matrices
\eq{
\h{\pi}(a)= \mb{Z{10}|Z{10}}{ 
\h{A} & 0 & 0 & 0 \\ 0 & \h{A} & 0 & 0 \\ \hline 
0 & 0 & \overline{\h{A}} & 0 \ru{2}{2} \\ 
0 & 0 & 0 & \overline{\h{A}} }~.~~ \label{pam}
   }
In terms of the decomposition $\C^{48}=(\ul{10} \op \ul5^* \op \ul1) 
\ot \C^3$ we have
\eq{
\h{A} = \diag \big( \pi_{10}(a) \ot \one_3~,~ \overline{\pi_5(a)} \ot \one_3~,~ 
0_3 \big) ~.~~	 \label{embA}
   }
Here, $\pi_5(a)=a$ is the adjoint representation $\ul{24}$ of $\su5$ 
and $\pi_{10}(a)$ the embedding of $\ul{24}$ into $\End(\ul{10})=\ul1 \op 
\ul{24} \op \ul{75}\,.$ The fact that the $\su5$ representations are tensorized 
by $\one_3$ means that the gauge group does not distinguish between the three 
generations of fermions. 

The mass matrix $\M$ of the L--cycle consists of two different contributions. 
The first one is diagonal and the other one off--diagonal in the sense of the 
indicated decomposition into two by two blocks in \rf[pam]:
\eq{
\M= \mb{Z{12}|Z{12}}{
0 & \M_i & \M_f & 0 \\ \M_i^* & 0 & 0 & \M_f \\ \hline 
\M_f^* & 0 & 0 & \overline{\M_i} \\ 0 & \M_f^* & \M_i^T & 0 }~.~~ 
   }
The $48 \times 48$--matrix $\M_f=\M_f^T$ is the fermionic mass matrix. A 
convenient picture is to imagine the two--two structure as the left--right 
decomposition. Since mass terms exchange left and right fermions, they must 
stand in the off--diagonal blocks. With this picture in mind it is not 
difficult to assign 
the $3 \times 3$--fermion mass matrices $M_u,M_d,M_e,M_n,M_N$ to the 
$16 \times 16$--block matrix $\M_f\,.$ Here, $M_u$ is the mass matrix for the 
$(u,c,t)$--quark sector, $M_d$ the mass matrix for the 
$(d,s,b)$--quark sector and $M_e$ the mass matrix for the 
$(e,\mu,\tau)$--lepton sector. Moreover, $M_n$ and $M_N$ are Dirac and Majorana 
mass matrices for the $(\nu_e,\nu_{\mu},\nu_{\tau})$--neutrino sector. 
These mass matrices include the fermion masses and generalized 
Kobayashi--Maskawa mixing angles. Mathematically, the sites where these 
generation matrices occur in $\M_f$ coincide with a combination of the 
representations $\ul5\,,$ $\ul{45}$ and $\ul{50}$ of $\su5\,.$ The relevant 
decomposition rules of tensor products are
\eq{
\Hom({\ul{10}}^*,\ul{10}) = \ul5^* \op \ul{45} \op \ul{50}~, \quad
\Hom(\ul5, \ul{10}) = \ul5 \op \ul{45}^* \;, \quad
\Hom(\ul1,{\ul5}^*) = \ul5^* \;.
   }
Let $n,n',m'$ be appropriate elements of $\ul5,\ul{45}^*,\ul{50}\,,$ in this 
order. Then one has 
\eq{
\M_f := \mbox{\small{$ \mb{ccc}{
{}~\iu \pi_{10,10}(n) \ot M_d + \iu m' \ot M_N & {}~~\iu \pi_{10,5} (n) \ot \Mu 
+ \iu n' \ot \Mn & 0 \\ 
{}~ \iu \pi_{10,5}(n)^T \ot \Mu^T + \iu n'{}^T \ot \Mn^T & 
0 & \iu \pi_{5,1}(n) \ot M_e ~{}  \\  
0 & \iu \pi_{5,1}(n)^T \ot M_e^T & 0 } $}}\,,  \label{embF}
  }
where $\pi_{10,10}(n)$ is the embedding of $n \in \ul5$ into 
$\Hom(\ul{10}^*,\ul{10})\,,$ $\pi_{10,5}(n)$ the embedding of $n$ into 
$\Hom(\ul5,\ul{10})$ and $\pi_{5,1}(n)$ the embedding of $n$ into 
$\Hom(\ul1,\ul5^*)\,.$ Moreover, 
\al{
\Mu &= \tfrac{1}{4} (3 M_u + M_n)~,& \Mn &= \tfrac{1}{4} (M_u - M_n)~.
  }
The block diagonal part $\M_i$ of $\M$ couples left--left and right--right 
sectors. Thus, it has no interpretation as fermion masses. It is responsible 
for the desired spontaneous symmetry breaking pattern from $\su5 \op \u1$ to 
$\su3 \op \su2 \op \u1 \op \u1\,,$ see item~\ref{sybr4}) at the very beginning. 
The non--Abelian part of $\su5 \op \u1$ commuting with $\M_i$ must coincide 
with the non--Abelian part of the standard model Lie algebra. In terms of 
the decomposition
\eq{
\su5 = \mb{E{15}|E{15}}{ \su3 & \cdot \\ \hline \cdot & \su2 }
   }
we put 
\eq{
m=\iu \,\diag(-\tfrac{2}{5}, -\tfrac{2}{5}, -\tfrac{2}{5}, 
\tfrac{3}{5}, \tfrac{3}{5} ) \in \su5~.~~
   }
With this notation, the desired symmetry breaking pattern is achieved for 
\eq{
\M_i := \diag( \iu \pi_{10}(m) \ot M_{10}~,~ 
\overline{ -\iu \pi_5(m) \ot M_5} ~,~ 0_3 ) ~,~~ \label{embJ}
  }
where $M_{10}$ and $M_5$ are arbitrary $3 \times 3$--matrices. In contrast to 
the parameters entering $M_f$ we have no experimental hints how to choose 
$M_{10}$ and $M_5$ except that their norm must be very large. Namely, in 
the flipped $\SU5 \times \U1$--GUT there occur interactions 
which lead to proton decay. It turns out that the lifetime predicted for the 
proton depends on 
$\tr(M_{10} M_{10}^* + M_5 M_5^*)\,.$ The larger the trace (in units of $m_t$), 
the larger is the lifetime of the proton. It is essential that the matrices 
$M_{u,d,e,n,N}$ and $M_{10,5}$ are generically chosen, because 
otherwise there would be unwanted contributions from the extension \rf[cpl].
Finally, the grading operator is 
\eq{
\h{\Gamma}= \diag \big(\,-\one_{48}\,,\, \one_{48}\,,\, \one_{48} \,,\, 
- \one_{48}\, \big)~.
   }

\subsubsection{Remarks on the Construction}

To this L--cycle we apply our formalism, which performs the following job: 
First, it extends the matrix $a \in \su5$ to a $\su5$--gauge field $A\,.$ 
This step is obvious, because we have $A \in \P{1}=\D{1}\,.$ Second, a rather 
long calculation reveals that those local elements of $\H{1}$ that are not 
already contained in $\P{1}$ are $\u1$--gauge fields $A''\,.$ The 
representations $\pi$ of $A$ and $A''$ on the fermionic Hilbert space are fixed 
by the formalism. In the notation of \rf[pam] they are given by 
\eqa{rcl}{
\pi (A) &=& \diag\big( \t{A}, \t{A}, \gamma_C \overline{\t{A}} \gamma_C, 
\gamma_C \overline{\t{A}} \gamma_C \big)~, \npb \\ 
\t{A} &=& \diag \big( \pi_{10}(A) \ot \one_3~,~ 
\gamma_C \overline{\pi_5(A)} \gamma_C \ot \one_3~,~ 0_3~ \big)	~,  
\label{embAp} \yn \\
\pi (A'') &=& \diag\big( \t{A}'', \t{A}'', \gamma_C \overline{\t{A}''} 
\gamma_C, \gamma_C \overline{\t{A}''} \gamma_C \big)~, \npb \\ 
\t{A}'' &=& \diag \big( {-} \tfrac{1}{2} A'' \one_{10} \ot \one_3~,~ 
{-} \tfrac{3}{2} \gamma_C \overline{A''} \gamma_C \one_5 \ot \one_3~,~ 
{-} \tfrac{5}{2} A'' \ot \one_3 \big)  ~, \quad{} \label{embApp} \yn
   }
where $\gamma_C$ is the complex conjugation matrix: $\g[\mu] = \gamma_C 
\overline{\g[\mu]} \gamma_C\,,$ $(\gamma_C)^2= \pm \one_4\,.$ Third, the 
formalism extends the matrices 

$m$ to a $\ul{24}$--Higgs multiplet $\ittJ=\itJ+m\,,$ 

$n$ to a complex $\ul5$--Higgs multiplet $\ittF=\itF+n\,,$ 

$n'$ to a complex $\ul{45}^*$--Higgs multiplet $\ittY=\itY+n'\,,$ 

$m'$ to a complex $\ul{50}$--Higgs multiplet $\ittX=\itX+m'\,.$ 
\\
This is an immediate consequence of the fact that $m,n,n',m'$ belong to 
irreducible representations. Thus, the formalism generates the complete bosonic 
configuration space of the flipped $\SU5 \times \U1$--model out of the given 
L--cycle. Totally, there are 224~Higgs fields and 25~gauge bosons. The 
connection form has the structure 
\eqa{l}{
\rho= \yn \npb \\
\mb[\!]{cccc}{
\t{\pi}(A{+}A'') & \g \t{\pi}(\ittJ) & \t{\pi}(\ittF{+}\ittX{+}\ittY) & 0 \\
-(\g \t{\pi}(\ittJ))^* & \t{\pi}(A{+}A'') & 0 & 
\g \t{\pi}(\ittF{+}\ittX{+}\ittY) \\ 
-(\g \t{\pi}(\ittF{+}\ittX{+}\ittY))^* & 0 &
- \gamma_C \overline{ (\t{\pi}(A {+}A''))} \gamma_C & 
\overline{ \g \t{\pi}(\ittJ)} \\
0 & -(\g \t{\pi}(\ittF{+}\ittX{+}\ittY))^* & 
- (\overline{\g \t{\pi}(\ittJ)})^* & -\gamma_C \overline{ (\t{\pi}(A{+}A''))} 
\gamma_C }. 
   }
Here, we have denoted by $\t{\pi}$ the embeddings \rf[embAp] and \rf[embApp] of 
the gauge fields $A$ and $A''\,,$ the embedding \rf[embF] of the Higgs 
multiplets $\ittF\,,\ \ittY$ and $\ittX$ and the embedding \rf[embJ] of the 
Higgs multiplet $\ittJ$ into $\mat{48}$ each. Thus, Yang--Mills and Higgs 
fields are treated in a unified way. Since the embeddings \rf[embF] and 
\rf[embJ] include the matrices $M_{u,d,e,n,N}$ and $M_{10,5}\,,$ the bosonic 
masses will depend on the fermion masses and the parameters of $M_{10,5}\,.$ 

The bosonic Lagrangian contains the usual Yang--Mills Lagrangian, the 
covariant derivatives of the Higgs fields and the Higgs potential. The Higgs 
potential is very complicated as a fourth order polynomial in 224 
variables. All gauge invariant combinations of 
\eq{
\pi_{10}(\ittJ)\,,\, \pi_{5}(\ittJ)\,,\, \pi_{10,10}(\ittF)\,,\, 
\pi_{10,5}(\ittF)\,,\,\pi_{5,1}(\ittF)\,,\, \ittY\,,\, \pi_{10,10}(\ittY)\,,\, 
\ittX
   }
really do occur. A computation of the minimum of such a monster seems hopeless. 
However, we do not have to work. The minimum is simply given by 
\al{
\ittJ &=m~,& \ittF &=n~,& \ittY &=n' & \ittX &=m' ~. \label{jmm}
   }
This is a general feature of both non--commutative and non--associative 
geometry; the Higgs fields occur already in the broken phase. Just to give an 
impression of the power of our approach we list few examples of occurring 
contributions to the Higgs potential. Let 
\eqa{rclrcl}{
V_1 &=& \ittJ^2 - \tfrac{1}{5} \tr(\ittJ^2) \one_5 - \tfrac{1}{5} \iu \ittJ ~,
\qquad{} & 
V_2 &=& ( \ittY \ittY^* )' + \tfrac{8}{3} \iu \ittJ - \ittF^* \ittF 
+ \tfrac{1}{5} \tr(\ittF^* \ittF) \one_5 ~, \npb \\
V_3 &=& \mc{4}{l}{ \ittY^* \itY - \tfrac{1}{5} \tr (\ittY^* \itY) \one_5 
+ 8 \iu \ittJ + 9 \ittF^* \itF - \tfrac{9}{5} \tr(\itF^* \itF) \one_5 ~,} 
\npb \\
V_4 &=& \mc{4}{l}{ \itY^* \pi_{10,5}(\ittF) + \pi_{10,5}(\ittF)^* \itY 
- 8 \iu \ittJ - 6 \itF^* \itF + \tfrac{6}{5} \tr(\itF^* \itF) \one_5 ~, } 
\label{vi} \npb \yn \\
V_5 &=& \itY^* \pi_{10,5}(\itF) - \pi_{10,5}(\itF)^* \itY ~, &
V_6 &=& (\ittX \ittX^*)' + \tfrac{1}{3} \iu \ittJ~.
    }
Here, $\iu Y'$ denotes the $\ul{24}$--component of the $10 \times 10$--matrix 
$\iu Y\,.$ Then, 
\eq{
\tsum_{i,j=1}^6 \mu_{ij} \tr(V_i V_j)  \label{vivj}
   }
is a typical contribution to the Higgs potential. If one came to the idea to 
change the relative coefficients a bit, say, to omit the linear terms in 
$V_i\,,$ then \rf[jmm] is no longer the minimum and one has to deal with the 
monster. At this point at the latest one realizes the advantage that 
non--associative geometry brings to gauge field theory. The linear terms in 
\rf[vi] arise from the part $\sigma(\omega^1)$ in equation \rf[pdw] for the 
differential. They lead to cubic terms in the Higgs potential, which must not 
be omitted! Principally, we have the freedom to choose the global parameters in 
the Higgs potential such as $\mu_{ij}$ in \rf[vivj] arbitrarily (but such that 
the Higgs potential remains positive definite). In the classical construction 
this freedom exists indeed, and that is the reason why one obtains no 
predictions for the masses of the Higgs fields. In our approach, 
also these global parameters are fixed. They are given by traces over certain 
combinations of the matrices $M_{u,d,e,n,N}$ and $M_{10,5}\,.$ Thus, if we fix 
the mass matrix $\M$ then all Higgs masses are determined on tree--level. 

In the flipped $\SU5 \times \U1$--model, the Lie subalgebra which leaves the 
vacuum \rf[jmm] invariant is $\CX \ot (\su3_C \op \u1_{EM})\,.$ The $\su3_C$ 
corresponds to the colour symmetry and the $\u1_{EM}$ to the symmetry 
generated by the electric charge of the particles. The remaining 16 gauge 
degrees of freedom, corresponding to 
\eq[rem]{
\CX \ot ((\su5 \op \u1)/(\su3_C \op \u1_{EM})) ~,
      }
are used to gauge away 16 Higgs fields, twelve of the 
$\ul{24}$--representation, three of the $\ul5$--representation and one of the 
$\ul{50}$--representation. This in turn gives a mass to the 16 former 
gauge bosons corresponding to \rf[rem]. These are the $W^\pm$ and $Z$ bosons, 
an additional neutral heavy gauge boson $Z'$ and the twelve leptoquarks $X$ 
and $Y$ (six each). Thus, there remain 208 Higgs fields
\eq{
\psi_1',\psi_2',\psi_3',\psi_0,\psi_1 ,\dots, \psi_8,\phi_0',\phi_1,\dots, 
\phi_6, \upsilon_0',\upsilon_1, \dots, \upsilon_{89},\xi_0,\xi_1, \dots, 
\xi_{98}\,,
   }
whose masses are obtained by diagonalization of the bilinear 
terms of the Higgs potential. These bilinear terms to select is still a 
tedious procedure (without computer algebra it is almost impossible to 
avoid errors). 
 
\subsubsection{The $\SU5$--Grand Unification Model}

If we omit ad hoc the $\u1$--gauge field $A''$ and put $M_N$ equal to zero, we 
can ``derive'' the $\SU5$--grand unification model out of the flipped 
$\SU5 \times \U1$--model. This derivation violates the principles of 
non--associative geometry. However, if we do not perform the extension 
\rf[cpl], then the $\SU5$--model is obtained from the same L--cycle introduced 
above, after renaming $M_u \leftrightarrow M_d\,,$ $M_n \mapsto M_e$ and 
$M_e \mapsto M_{\nu}\,.$ If one omits the $\ul{5}$--representations and the 
matrix $M_{\nu}$ then one gets a model without right neutrinos. 

\subsubsection{Physical Results from the Grand Unification Model}
 
We present the final results (on tree--level) for the flipped 
$\SU5 \times \U1$--grand unification model in Table~\ref{tab0}. 
\begin{table}[p]
\vs*{-4ex}
{}\hfill$
\begin{array}{|E{34.5}|E{34.5}|E{1.8}|E{34.5}|E{34.5}|}
\cline{1-2} \cline{4-5}
\mbox{ Particle } & \mbox{ Mass} && 
\mbox{ Particle } & \mbox{ Mass} 
\\ 
\cline{1-2} \cline{4-5} 
\mc{5}{c}{ \mbox{ \textbf{1.} The completely neutral Higgs fields: \ru{3}{0} }} 
\\ \cline{1-2} \cline{4-5}
\phi_0'   & (0 \dots 1.45) \,m_t  &&
\xi_0	 & (\sqrt{\tfrac{1}{60}} \dots \sqrt{\tfrac{7}{4}}) m_N \\
\upsilon_0'   & \lambda m_t                 && 
\upsilon_{45} & \th \sqrt{3} \lambda m_t    \\
\psi_0	  & \sqrt{\tfrac{2}{5}} m_N &&
\psi_3'   & (0 \dots \tfrac{1}{12} \sqrt{\tfrac{11}{3}}) \tfrac{m_N^2}{M}
\\ 
\cline{1-2} \cline{4-5} 
\mc{5}{c}{ \mbox{ \textbf{2.} The colour--neutral Higgs fields of charge 
$\mp 1\,$: \ru{3}{0} }} 
\\ \cline{1-2} \cline{4-5} 
\tfrac{1}{\sqrt{2}} (\upsilon_{18} \pm \iu \upsilon_{63} ) & 
\th \sqrt{3} \lambda m_t && 
\tfrac{1}{\sqrt{2}} (\psi_1 \pm \iu \psi_2 ) &	    
	  (0 \dots \tfrac{1}{12} \sqrt{\tfrac{11}{3}}) \tfrac{m_N^2}{M}
\\ 
\cline{1-2} \cline{4-5} 
\mc{5}{c}{ \mbox{ \textbf{3.} The neutral Higgs fields, for $i=0, \dots ,7\,$: 
\ru{3}{0} }} 
\\ \cline{1-2} \cline{4-5} 
\psi_{i+1} & (0 \dots \tfrac{1}{12} \sqrt{\tfrac{11}{3}}) \tfrac{m_N^2}{M} & && 
\\
\upsilon_{i+1} & (\lambda \dots \lambda {+} \check{\lambda}) m_n &&
\upsilon_{i+45} & (\lambda \dots \lambda {+} \check{\lambda}) m_n \\
\xi_{i+32} &  3 M &&
\xi_{i+81} &  3 M 
\\ 
\cline{1-2} \cline{4-5} 
\mc{5}{c}{ \mbox{ \textbf{4.} The Higgs fields of charge $\mp 1\,,$ for 
$i=0 \dots 7\,$: \ru{3}{0} }} 
\\ \cline{1-2} \cline{4-5} 
\tfrac{1}{\sqrt{2}} (\upsilon_{19+i} \pm \iu \upsilon_{64+i} ) & 
       (\lambda \dots \lambda {+} \check{\lambda}) m_n &&
\tfrac{1}{\sqrt{2}} (\xi_{25+i} \pm \iu \xi_{74+i} ) & 3 M 
\\ 
\cline{1-2} \cline{4-5} 
\mc{5}{c}{ \mbox{ \textbf{5.} The Higgs fields of charge $\mp \tfrac{1}{3}\,,$ 
for $i=0,1,2$ and $j=0,\dots,5\,$: \ru{3}{0} }} 
\\ \cline{1-2} \cline{4-5}
\tfrac{1}{\sqrt{2}} (\phi_{1+i} \pm \iu \phi_{3+i} )	& M &&
\tfrac{1}{\sqrt{2}} (\upsilon_{9+i} \pm \iu \upsilon_{54 +i} )	& M \\
\tfrac{1}{\sqrt{2}} (\upsilon_{12+i} \pm \iu \upsilon_{57 +i} ) & M &&
\tfrac{1}{\sqrt{2}} (\upsilon_{39+i} \pm \iu \upsilon_{84 +i} ) & 2 M \\
\tfrac{1}{\sqrt{2}} (\xi_{44+i} \pm \iu \upsilon_{93 +i} ) & M &&
\tfrac{1}{\sqrt{2}} (\xi_{47+i} \pm \iu \upsilon_{96+i} ) & 2 M \\
\tfrac{1}{\sqrt{2}} (\xi_{19+j} \pm \iu \upsilon_{68+j} ) & 2 M &&
\tfrac{1}{\sqrt{2}} (\upsilon_{30+j} \pm \iu \upsilon_{75+j} ) & M 
\\ 
\cline{1-2} \cline{4-5}
\mc{5}{c}{ \mbox{ \textbf{6.} The Higgs fields of charge $\pm \tfrac{2}{3}\,,$ 
for $i=0,1,2$ and $j=0,\dots,5\,$: \ru{3}{0} }} 
\\ \cline{1-2} \cline{4-5}
\tfrac{1}{\sqrt{2}} (\upsilon_{15+i} \pm \iu \upsilon_{60 +i} )  & M &&
\tfrac{1}{\sqrt{2}} (\upsilon_{36+i} \pm \iu \upsilon_{81 +i} ) & 2 M \\
\tfrac{1}{\sqrt{2}} (\upsilon_{42+i} \pm \iu \upsilon_{87 +i} ) & M &&
\tfrac{1}{\sqrt{2}} (\xi_{41+i} \pm \iu \upsilon_{90 +i} ) & M \\
\tfrac{1}{\sqrt{2}} (\xi_{7 +j} \pm \iu \xi_{56 +j} ) & 2 M &&
\tfrac{1}{\sqrt{2}} (\xi_{13+j} \pm \iu \xi_{62 +j} ) & 4 M 
\\ 
\cline{1-2} \cline{4-5}
\mc{5}{c}{ \mbox{ \textbf{7.} The Higgs fields of charge $\mp \tfrac{4}{3}\,,$ 
for $i=0,1,2$ and $j=0,\dots,5\,$: \ru{3}{0} }} 
\\ \cline{1-2} \cline{4-5}
\tfrac{1}{\sqrt{2}} (\upsilon_{27+i} \pm \iu \upsilon_{72+i} )	& M &&
\tfrac{1}{\sqrt{2}} (\xi_{1+j} \pm \iu \upsilon_{50+j} )  & 2 M 
\\ 
\cline{1-2} \cline{4-5}
\mc{5}{c}{ \mbox{ \textbf{8.} The neutral massive gauge fields: \ru{3}{0} }} 
\\ \cline{1-2} \cline{4-5} 
Z & \sqrt{\tfrac{2}{5}} \,m_t && 
Z' & \th \sqrt{\tfrac{5}{3}} m_N 
\\ 
\cline{1-2} \cline{4-5}
\mc{5}{c}{ \mbox{ \textbf{9.} The massive gauge fields of charge $\pm 1\,$: 
\ru{3}{0} }}
\\ \cline{1-2} \cline{4-5} 
\tfrac{1}{\sqrt{2}} (W_1 \mp \iu W_2) & \th m_t &&
\mc{2}{|c|}{ \mbox{Weinberg angle: } \sin^2 \theta_W = \tfrac{3}{8} } 
\\ 
\cline{1-2} \cline{4-5}
\mc{5}{c}{ \mbox{ \textbf{10.} The leptoquarks leading to proton decay, 
for $i=0,1,2\,$: \ru{3}{0} }}
\\ \cline{1-2} \cline{4-5}
\tfrac{1}{\sqrt{2}} (X_{1+i} \mp \iu X_{3+i} ) & M && 
\mc{2}{|c|}{ \mbox{charge: } \mp \tfrac{1}{3} } \\
\tfrac{1}{\sqrt{2}} (Y_{1+i} \mp \iu Y_{3+i} ) & M && 
\mc{2}{|c|}{ \mbox{charge: } \pm \tfrac{2}{3} }
\\ \cline{1-2} \cline{4-5}
\end{array}$
\vs*{-1.2ex}
\caption{The particle masses for the $\SU5 \times \U1$--model } \label{tab0}
\vs*{-5.6ex}
\end{table}%
In this table, we denote by $m_t$ and $m_b$ the masses of the top quark and by 
$m_n$ and $m_N$ the mass scales of the Dirac and Majorana masses for the 
neutrinos, respectively. The masses in Table~\ref{tab0} are correct for 
\eq{
m_n,m_b < m_t \ll \lambda m_t, (\lambda+\check{\lambda}) m_n < M,m_N~,~~
   }
which is physically plausible. The parameter $M \gg m_t$ is the grand 
unification scale. Moreover, we assume that the Majorana mass of the right 
neutrinos is of the same order of magnitude as $M\,.$ The parameters 
$M,\lambda,\check{\lambda}$ are certain combinations of the unknown parameters 
of the matrices $M_{10}$ and $M_5\,.$ For generic matrices $M_{10}$ and 
$M_5\,,$ the masses $\lambda m_t$ and $(\lambda \dots \lambda 
+ \check{\lambda}) m_n$ are not significantly smaller than $M$ and $m_N\,.$ 
Let us comment on some observations:  %(
\settowidth\labelwidth{4)}%
	    \leftmargini\labelwidth
            \advance\leftmargini\labelsep
\begin{enumerate}
\parskip0pt

\item
There occur three mass scales in the flipped $\SU5 \times \U1$--model: The mass 
scale of the fermions determined by $m_t\,,$ the grand unification scale $M$ 
and an intermediate scale determined by $\lambda m_t$ and 
$(\lambda \dots \check{\lambda}) m_n\,.$ All particles with fractional--valued 
electric charge, which therefore lead to proton decay, have a mass of the 
order $M\,.$

\item
There exists precisely one light Higgs 
field $\phi_0'\,,$ whose upper bound for the mass is independent of the grand 
unification matrices $M_{10}$ and $M_5\,.$ The reason that only an upper bound 
can be given is the incomplete knowledge of the input parameters. 
The Higgs field $\phi_0'$ is a certain linear combination 
of neutral Higgs fields of the $\ul5$--representation and the 
$\ul{45}^*$--representation\footnote{This shows impressively that the 
$\ul{45}$--representation, which is absent in the NCG--formulations 
\cite{cff1, cff2} of the $\SU5$--GUT, is an essential part of our model.}. 
It has precisely the same properties as the standard model Higgs field. 

\item
The predictions for the $\SU5$--model are qualitatively the same, except 
that the gauge field $Z'$ and all Higgs fields $\xi_i$ are absent. Moreover, 
the electric charges of certain Higgs fields are modified. 

\item
The standard model is in perfect agreement with experiment. However, our 
results show that the low energy sector of both the $\SU5 \times \U1$ and 
$\SU5$ GUT's is identical with the standard model. This 
means that it is not possible to decide by means of present energy experiments 
which of the three models is correct. One essential advantage of the grand 
unification models is that they explain why proton and electron have up to the 
sign the same electric charge. On the other hand, the proton is not a stable 
particle in grand unified models. Concerning this question, the 
$\SU5 \times \U1$--model is favoured over the $\SU5$--model, because it yields 
a larger lifetime for the proton \cite{dkn}. 

\end{enumerate}

\noindent
We see that non--associative geometry has the flexibility to describe grand 
unification models.

\end{document}